# X-Ray microtomography of mercury intruded compacted clay: an insight into the geometry of macropores

Shengyang Yuan[1], Xianfeng Liu[2], Yongxing Wang[3],
Pierre Delage[4], Patrick Aimedieu[5], Olivier Buzzi[6],

[1,2,3] Key Laboratory of High-speed Railway Engineering of Ministry of Education, School of Civil Engineering, Southwest Jiaotong University, Chengdu 610031, China.

[1,2,6] Priority Research Centre for Geotechnical Science and Engineering, The University of Newcastle, University of drive, Callaghan, NSW, 2308, Australia.

[2] Xinjiang Institute of Engineering, Urumchi, Xinjiang 860023, China.

[4,5] École nationale des ponts et chaussées, Institut Polytechnique de Paris, Laboratoire Navier, 6–8 av. Blaise Pascal, Marne-la-Vallée, France.

**Corresponding author** Dr. Xianfeng Liu, Xianfeng.liu@swjtu.edu.cn

# 1. Abstract:

Soil properties, such as wetting collapse behavior and permeability, are strongly correlated to the soil microstructure. To date, several techniques including mercury intrusion porosimetry (MIP), can be used to characterize the microstructure of soil, but all techniques have their own limitations. In this study, the features of mercury that penetrated and has been entrapped in the pore network of the specimens through MIP testing were investigated by X-Ray microtomography (X-μCT), in order to give an insight into the geometry of macropores and possible ink-bottle geometry. Two conditions of water content and density were selected for the compacted Maryland clay. The distribution and geometry features of mercury entrapped in the microstructure after MIP were characterized and pore size distributions were also reconstructed. The results suggest that, for the two conditions studied in this paper, macropores were evenly distributed within the specimens, and most of them with a non-spherical shape, and with aspect ratio (ratio between the maximum and minimum thickness along a given segment) smaller than three. Different dominant entrance pore size of macropore was obtained from MIP and X-μCT, due to the specific experimental protocol used in tests and the effect of ink-bottle geometry. Only the large pore bodies with high aspect ratio were imaged in X-μCT, due to the extrusion of mercury during the process of depressurization and subsequent sample preparation for X- μCT. But entire pore space was accessible in MIP. The difference in dominant entrance pore size was more significant for specimens with lower void ratio due to a more pronounced aspect ratio.

**Key words**: Microstructure, Expansive soils, Mercury, Macropores, Mercury intrusion porosimetry, X-ray tomography

# 2. Introduction

Soil microstructure strongly contributes to the mechanical and hydraulic properties of soils such as permeability (Garcia-Bengochea et al., 1979; Lapierre et al., 1990; Yuan et al., 2019b), compressibility (Delage, 2010; Delage and Lefebvre, 1984), water retention (Romero et al., 2011; Seiphoori et al., 2014; Yuan et al., 2019b) and strength (Liu et al., 2016; Ng et al., 2020). As pointed out by, among others, Yuan et al. (2019a, 2019b, 2021a), wetting collapse behaviour, permeability and swelling behaviour are strongly affected by the macropores of soil. The pore structure of a porous medium is commonly characterised via scanning electron microscopy (SEM) (e.g. Collins and Mcgown, 1974; Delage, 2010; Monroy et al., 2010; Seiphoori et al., 2014), mercury intrusion porosimetry (MIP) (Ahmed et al., 1974; Delage et al., 1996; Diamond, 1970; Monroy et al., 2010, Yuan et al., 2020), nitrogen adsorption (Cases et al., 1997; Otalvaro et al., 2016) and X-ray computed microtomography (X-μCT) (Bruchon et al., 2013; Cnudde and Boone, 2013). Each technique possesses its own limitations. For example, SEM can only characterize 2D microstructural features of specimens (Romero and Simms, 2008). Nitrogen adsorption can only be used to characterize pore diameters up to 50 nm because of the size of nitrogen molecules (Cases et al., 1997). As for MIP, although it can provide a pore size distribution (PSD) ranging from few nanometers to several hundreds of micrometers, the so-called "ink-bottle geometry" which refers to a specific pore shape where large cavities are connected by narrow connections (Giesche, 2006; Moro and Böhni, 2002; Ravikovitch and Neimark, 2002), and ink-bottle ratio or aspect ratio is defined as the ratio between pore body and pore throat, can lead to an overestimation of the volume of smaller pores. Note that, in this case, further precision can be obtained by monitoring the mercury pressure release curve, so as to determine the entrapped porosity that corresponds to the volume of mercury blocked in ink-bottle shaped (or constricted) pores. In this regard, Delage et al. (1984) have shown that the inter-aggregate pores of sensitive clays were entrapping the mercury while releasing the mercury pressure, and that the ink-bottle effect was due to the constriction caused by the inter-aggregate clayey bridges. Conversely, the intra-aggregate pores did not present any ink-bottle effect.

Nitrogen adsorption and MIP tests provide PSD of specimens but cannot provide 3D information about pore structures. X-μCT can be used to generate 3D images of specimen based on the variation of X-Ray attenuation within the specimen, which relates to the composition and density of specimen. Recently, it has been used to characterize the 3D morphology of particles of compacted bentonite/sand mixture (Saba et al., 2014), sands (Bruchon et al., 2013; Fonseca et al., 2012; Le et al., 2020) and rocks (Charalampidou et al., 2011; Sufian and Russell, 2013). X-μCT is particularly useful for the determination of pore body to pore throat aspect ratio (Andersson et al., 2018; Gerke et al., 2020). The accuracy of results based on X-μCT mainly depends on device, imaging settings, and image processing. Artifacts, like ring artifacts, intensity bias, streaking artifacts, should be removed first, then followed by denoising. Each grey scale image needs threshold to give a binary pore solid image, then segmentation is carried out to partition the image into multiple segments (Iassonov et al., 2009; Schluter et al., 2014). During the structural analysis, the size of pore body and pore throat size can be determined based on the maximum inscribed sphere method (Silin and Patzek, 2006; Tanino and Blunt, 2012). But all aforementioned pores are in a comparative large size, in order of few hundreds micrometers to few mimimeters (Andersson et al., 2018; Bruchon et al., 2013; Fonseca et al., 2012; Le et al., 2020; Sufian and Russell, 2013). For clayey soils, especially reactive clays, inter-aggregate pores are generally in order of micrometers (see Delage et al., 2006; Monroy et al., 2010; Romero, 2013; Yuan et al., 2021a). As a consequence, it is difficult to characterise a clay microstructure using X-μCT technique, as this is pushing the limits of the technique resolution. Only few studies has been carried out, such as Massat et al. (2016) characterized the evolution of inter-aggregate porosity along the specimen height of an expansive soil during hydration using X-μCT, but they did not characterize the features of pores (i.e. flatness, anisotropy).

In this study, MIP and X-μCT techniques were combined in order to characterize the macropore structure (macropores are typically considered to be pores having an entrance pore size in excess of 1 μm, Yuan et al., 2020) of compacted clayey soils. The idea was to detect and characterise the entrapped mercury present in the soil pores of a sample previously submitted to a MIP test by using X-μCT, taking advantage of the very high density of mercury compared to that of the soil particles. The geometry features of macropores for compacted expansive clays were studied, including the distribution, shape of these pores, and the ink-bottle geometry of the pore structure was evaluated. This study provided the first 3D insight into the pores characterised by mercury intrusion porosimetry.

## 3. Materials and methods

A natural expansive clay from the in-situ testing field of Maryland, Newcastle, Australia was used (Fityus et al., 2004). Maryland clay (referred to as ML clay) contains quartz (36.9% in mass), kaolinite (26.6% in mass), mica (17.4% in mass), inter-layered illite-smectite (10% in mass), plagioclase (5.8%), and K-feldspar (3.3%). The clay fraction (<2 μm) represents 71.8% in mass. The soil has a specific gravity equal to 2.65. The plastic and liquid limit are 24.1% and 69.8%, respectively. The cation exchange capacity, determined by Melton method (French standard-NF X 31-130, AFNOR 1999), is 14 mequiv./100 g. For more details about the physical properties of ML clay, refer to Fityus and Smith (2004) and Liu et al. (2016). The specimen preparation procedure used herein was similar to that followed in Yuan et al. (2016):

peds of oven dried (at 105 ºC for 12 hours) soil were crushed and particles retained on the 1.18 mm sieve were discarded. Roots and other visible organic matter were also removed. The soil powder was then hydrated with distilled water by hand spraying to a target water content and left to equilibrate in airtight containers for at least two days. The soil was then compacted at a constant displacement rate (1mm/min) to reach a target void ratio in dimension of 19 mm in height and 45 mm in diameter.

When selecting the initial conditions for this study, it was significant to consider the resolution of the tomograph (here 5 μm/voxel, where voxel includes the pixel value and thickness of slice) and the rule of thumb according to which a minimum of 2 to 3 voxels were required in order to give a better detection of a volume.

The publication by Yuan et al. (2021a) presents extensive results of microstructural analyses of ML clay compacted at void ratios ranging from 0.6 to ~1.3 and water contents ranging from ~12% to ~30%. The PSDs presented in Yuan et al. (2021a) suggest that, in order to obtain a significant amount of macropores in the size range of 10 to 15 μm, it is best to use a moisture content in the range ~13% to ~22% and a void ratio equal or larger than 0.8. Here, two initial conditions for compaction were selected (condition 1: initial water content, $w_o = 21\%$, and initial void ratio, $e_o =1$; condition 2: $w_o = 13.2\%$ and $e_o =0.8$). Corresponding PSDs were reproduced in Figure 1.

As shown in Figure 1, a clear bi-model structure was observed for both conditions. The entrance pore size for the peak corresponding to macropores is much larger for condition 1 than for condition 2 because of the higher void ratio and the higher water content. Indeed, macropores are reduced by compaction (i.e. lower void ratio, Monroy et al., 2010) and wetting of ML clay induces swelling of the aggregates and opening of the macropores (Yuan et al. 2021a).

Small prisms (of dimensions approximately equal to 1 cm × 0.5 cm × 0.5 cm) were cut with a wire saw from the compacted specimens and freeze-dried (as per Yuan et al., 2021b). Freeze-drying is a standard preparation technique to empty the pores of their water with minimal changes in the soil microstructure (Delage et al., 1982; Gillott, 1973; Tovey and Wong, 1973). The effect of the freeze-drying process on soil volume change has been studied by Zimmie and Almaleh (1976), using a kaolinite cube (7 mm on a side, initial water content ranging from 10% to 45%). The authors found that, with only about 4% of volume shrinkage, freeze drying induces less volume change than air drying and oven drying and preserve the soil microstructure. One inherent difficulty of performing tomographic tests on mercury intruded specimens is the very high density of mercury leading to significant beam attenuation. To mitigate this issue, the freeze-dried specimens were broken into smaller peds (in irregular shape with dimensions approximately equal to 0.5 cm × 0.25 cm × 0.25 cm) before performing the porosimetry and tomographic tests, with the direction of long axis corresponding to the vertical direction. A mass of about 0.7 g of soil was used for each MIP tests and 4 small specimens (2 for each set of initial conditions) were selected for the tomographic tests (see Table 1).

In this study, the MIP tests were conducted using an Autopore IV (Micromeritics), which has a maximum pressure is 240 MPa. For this study, a maximum injection pressure of 103 kPa and 138 kPa (corresponding to an entrance pore size about 15 μm and 10 μm) was applied for condition 1 and 2, respectively.

The values of maximum injection pressure were decided based upon two considerations: first, given the resolution of the tomograph used, there is no point injecting mercury into pores smaller than 10 to 15 μm as these are unlikely to be detected. Second, the mercury present in the intruded specimens should not be subjected to an increase in pressure when the specimens are removed from the porosimeter, as this could affect the mercury distribution inside the specimens. Consequently, the maximum pressure has to be larger than atmospheric pressure.

Table 1: Initial conditions of specimens used for X-μCT tests

| Conditions | Specimen ID | Volume of specimen, mm³ | Maximum injection pressure, kPa |
|---|---|---|---|
| Condition 1 | S1 | 31 | 103 |
| $w_o$=21%, $e_o$=1.0 | S2 | 19 | 103 |
| Condition 2 | S3 | 21 | 138 |
| $w_o$=13.2%, $e_o$=0.8 | S4 | 14 | 138 |

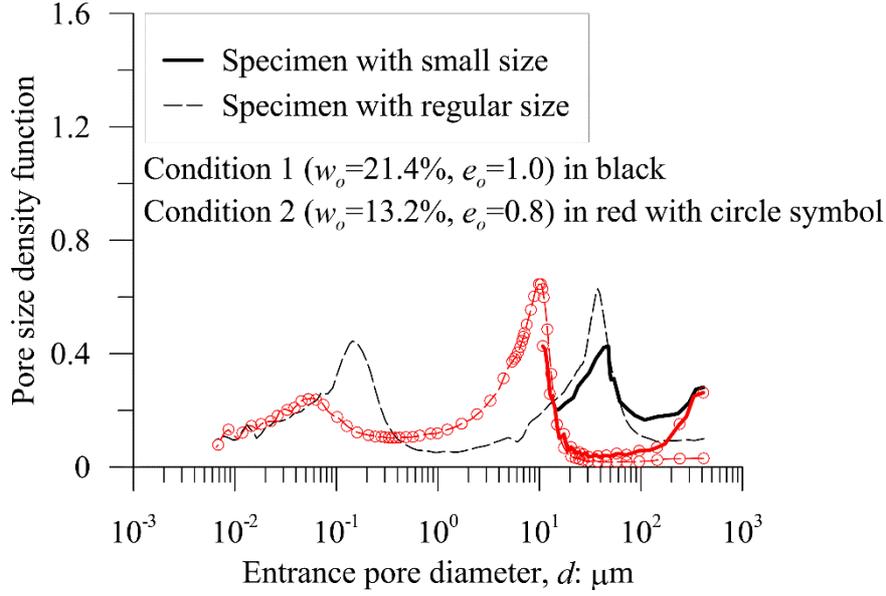

Figure 1: Pore size distribution of specimens for X-μCT and full mercury intrusion. Note condition 1 is in black colour, and condition 2 is in red colour. The dashed lines correspond to the full pore size distribution for specimen with regular size. While the bold continuous lines correspond to a partial pore size distribution obtained on the specimens to be subjected to X-μCT in small size. Voxel size on images obtained by X-μCT is equal to 5 μm (reproduced from Yuan et al., 2019a).

After the MIP tests, the mercury intruded specimens were recovered and sealed in epoxy resin to be visualized using tomography. The mercury intruded specimens were first glued to the end of a wooden toothpick stuck in plasticine, with the long axis in vertical direction. A plastic straw was then placed around the toothpick and the specimen before filling the straw with epoxy resin (see Figure S1 in Supplementary Martial, accessible online).

The microtomographic apparatus used in this study is located at the Navier institute, located in Ecole Nationale de Ponts et Chaussées, Champs-sur-Marne, France. It is an Ultratom, made by RX Solutions, which combines two sources (160kV and 230kV) which can be used separately, depending on the resolution needed. The source used for the present study was the 230kV (model Hamamatsu L10801), which allows obtaining 4 μm voxels at best. The X-ray detector was a Varian 2520 (pixels: 1840 × 1456). Acceleration voltage 200 kV, and electron current 25 μA were used without X-ray filter. Three frames were taken by imager per second. For every position of the rotation stage, 12 projections were taken and averaged to reduce noise. Total 1440 projections were used to cover 360º, and the scan duration for one specimen was around 100 min.

The 3D distribution of mercury in the specimens was reconstructed with the Avizo 2019.1 software (Avizo, 2018). A grey level (GL) value of an X-ray tomographic image is a function of the linear attenuation coefficient, which depends on X-ray energy, and material and its bulk density. The denser the material, the higher the GL. Analysing the images (with 16 bits per voxel) showed that mercury with a GL around 65500, soil particles with a GL around 22000 and epoxy with a GL around 18900 (Figure 2a). Given that the focus of this study is to analyse the shape of intruded mercury, a threshold GL has been applied to all images in order to clearly distinguish the mercury fraction from the non-mercury fraction, i.e. soil, air, resin lumped together.

In defining the GL threshold, it is relevant to bear in mind that underestimating the threshold will result in an overestimation of the mercury fraction, and vice-versa. As shown in Figure 2a, there is significance different between the GL of mercury and the rest parts (soil particles and epoxy), two clearly different intensity values can be obtained in the histogram of this figure, so during the image processing phase, the Otsu's threshold clustering algorithm was used (Otsu, 1979) combined to a manual adjustment of the GL threshold. Determining the appropriate value of threshold GL was done following the method proposed by Wong et al. (2006) and used by other researchers (e.g. Attari et al., 2016; Zeng et al., 2020). The method consists of plotting the area of segments against the value of GL threshold and using the intersection of asymptotes as the threshold (see Figure 2c, example for the slice shown in Figure 2a). As shown in Figure 2c, the grey level for the inflection point is around 26000, and the corresponding segmentation is shown in Figure 2b. This exercise was conducted for each specimen.

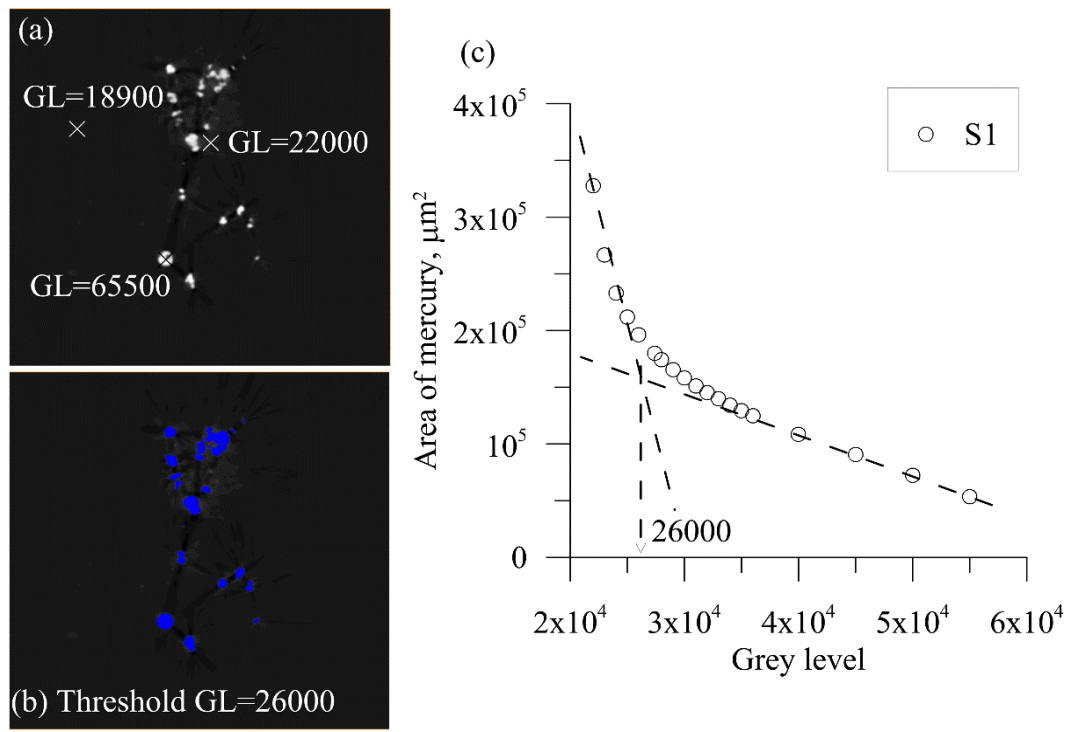

Figure 2: (a) view of mercury, resin and soil fractions with corresponding grey level (GL); (b) view of mercury fraction (in blue) at GL threshold 26000; (c) relationship between threshold grey level and area of mercury fraction on the processed image, called "area of mercury".

Once the mercury fraction was clearly identified, its 3D construction and analysis were undertaken (including label analysis and skeletonization process, Delerue et al., 1999) in order to obtain features such as orientation, thickness, flatness and anisotropy.

## 4. Results and discussion

### 4.1 Pore size distribution of specimens obtained from MIP tests

The soil specimens were intruded with mercury prior to conducting the X-μCT tests, resulting in partial pore size distributions that corresponds to the entrapped porosity affected by ink-bottle geometry, which were compared to the pore size distribution presented in Figure 1 and obtained from MIP test of specimens compacted under the same initial conditions. Overall, the partial PSDs are very similar to the full PSDs but for the fact that the partial PSDs contain much more large pores (size in excess of 100 μm), which is due the possible cracks created on the peds surface during the preparation process (breaking the soil specimen into small peds after freeze-drying) and the fact that the specimens to be tested in X-μCT consists of an assembly of small clay peds (volume in the order of 20 to 30 mm³, see Table 1) with large pores in between, but not within, peds. Note that it is expected that, during unloading, the mercury present in the superficial cracks would be completely extruded and, as a result, would not be visible in the X-μCT images.

Looking at the PSDs for condition 1 (Figure 1), the two peaks are almost aligned (dominant entrance pore sizes of 37.6 μm and 47.7 μm) but the frequency of large pores is higher for the full PSD, which is not explained and could simply be inherent variability. For condition 2 (Figure 1), except for the very large pores discussed above, the partial PSD follows the full PSD. The small pore size of the partial PSD is 10.8 μm.

### 4.2 Distribution of mercury in the soil specimens

Figure S2a and S2b present the reconstructed 3D shape of mercury for specimen S2 (condition 1) and S4 (condition 2), respectively. The basic shape of all mercury volumes can be approximated as near spherical shape or non-spherical shape (see Figure S1c). Note that, for each condition, all specimens tested have a mercury distribution similar to the respective distribution shown in Figure S2. It is here suggested that the mercury forms near spherical shapes when the volume of mercury is smaller than that of the pore and the mercury contracts back into a near sphere. In contrast, a non-spherical shape suggests that the mercury fills the majority of a pore and adopts most of its shape.

The orientation, barycentre position, flatness and anisotropy of each isolated volume of mercury was statistically studied. Flatness and anisotropy are defined from the eigenvalues of the covariance matrix for moments of inertia, which is similar to defining particle shapes from dimensions along the short, intermediate and long axis (Avizo, 2018). Flatness ($F$) and anisotropy ($A$) can be calculated based on the eigenvalue of covariance matrix for moments of inertia by following equation:

$$F = \lambda 1 \, / \, \lambda 2 \qquad \qquad \text{Eq. 1}$$

$$A = 1 - \lambda 1 \, / \, \lambda 3 \qquad \qquad \text{Eq. 2}$$

where $\lambda 1$, $\lambda 2$, and $\lambda 3$ are the smallest, medium, and largest eigenvalue of the covariance matrix. More details for the covariance matrix for moments of inertia please see in supplementary material.

The shape of mercury volume can be first study by anisotropy. $A$ around 0 indicates a near spherical shape. As shown in Figure S3, the cumulative frequency of $A$ for the four specimens are almost the same, with around 90% of mercury with $A$ larger than 0.4.

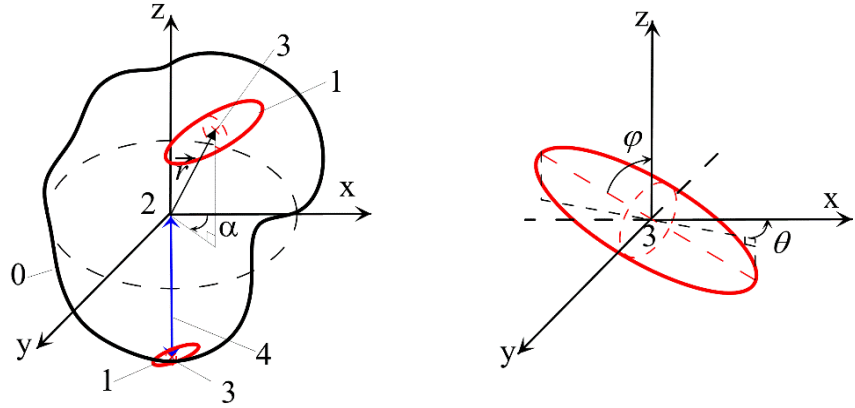

Figure 3: Schematical representation of relative position of an isolated mercury volume with respect to soil specimen; and orientation of isolated mercury volume. Note that vector $\vec{r}$ goes from the barycentre of soil specimen to that of mercury, and the length of $\vec{r}$ is denoted by r; 0: soil specimen; 1: mercury volume; 2: barycentre of the soil specimen; 3: barycentre of isolated mercury volume; 4: longest distance from the barycentre of the soil specimen to the barycentre of isolated mercury volume (noted rmax); f is the angle between largest eigenvector and z-axis; q is the angle between the project of largest eigenvector in x-y plane and x-axis; azimuthal angle a about x-y plane, starts from x-axis and rotates clockwise to the projection of r in x-y plane.

In order to describe the position of each isolated mercury volume (near spherical or non-spherical shape, capture through the X-µCT), the vector $\vec{r}$ (length of vector $\vec{r}$ is equal to $r$) and azimuthal angle $\alpha$ are introduced as per Figure 3. The z-axis shown in Figure 3 is consistent to the vertical direction shown in Figure S1. How deep did mercury penetrate into specimen was characterized by normalized distance, $r/r_{max}$.

Figure 4 shows the relationship between azimuthal angle $\alpha$ and $(r/r_{max})^{1.5}$ in a polar plot. $(r/r_{max})^{1.5}$ is used to ensure that the uniformly distributed data in 3D also looks uniformly distributed in 2D polar plot. Figure 4a and Figure 4b correspond to conditions 1 and 2, respectively. When analysing the figures, the following needs to be considered: (1) the smaller of distance ratio $r/r_{max}$, the deeper the mercury has penetrated into the soil specimen and (2) the azimuthal angle $\alpha$ represents the isotropy of penetration of mercury in the soil specimen and will reveal the presence or absence of preferential direction of penetration. It can be seen that the distribution of mercury volumes is similar for the specimens with same condition (Figure 4), and for both conditions. Figure 4 shows that mercury penetrates in the specimens from all directions and distributes more or less uniformly in specimens, except for condition 2, very little mercury penetrates further than $(r/r_{max})^{1.5}=0.15$ (for $r/r_{max}$ around 0.28). It is mainly due to the density for condition 2 is slightly higher than that of condition 1.

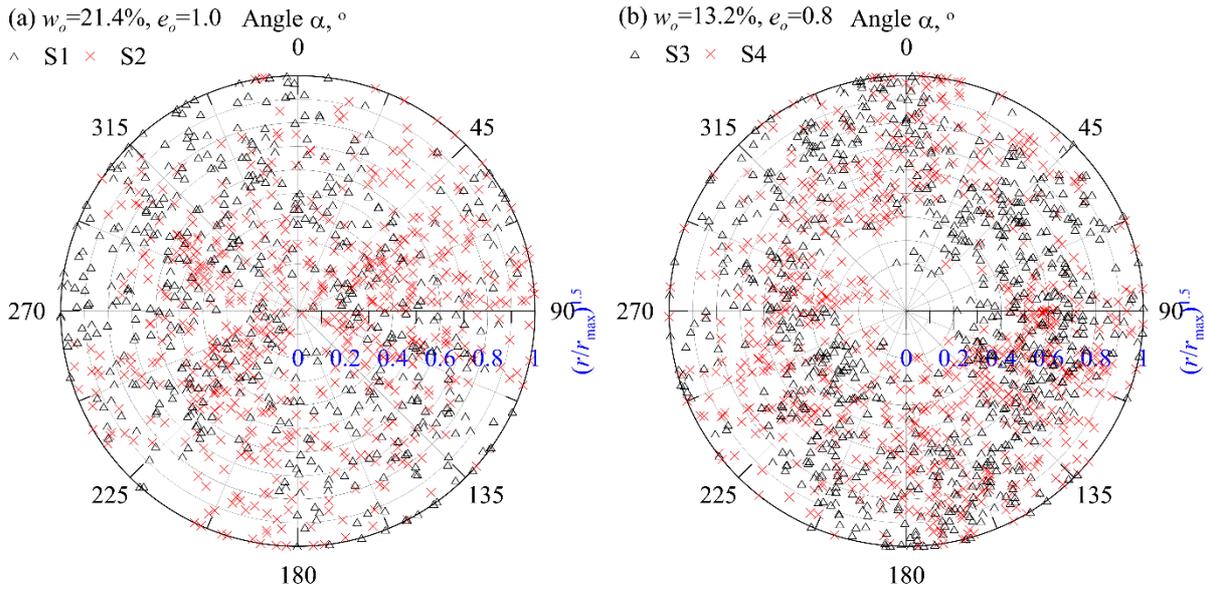

Figure 4: Polar plot of azimuthal angle $\alpha$ and $(r/r_{max})^{1.5}$: a) specimens S1, S2 (condition 1, $w_o$=21%, $e_o$=1.0); b) specimens S3, S4 (condition 2, $w_o$=13.2%, $e_o$=0.8).

Figure 5 now shows the cumulative distribution of distance ratio $r/r_{max}$ (Figure 5a) and azimuthal angle $\alpha$ (Figure 5b) for the 4 specimens. The theoretical distribution of $r/r_{max}$ and $\alpha$ of an ideal specimen is also presented in Figure 5, with the corresponding theoretical equations.

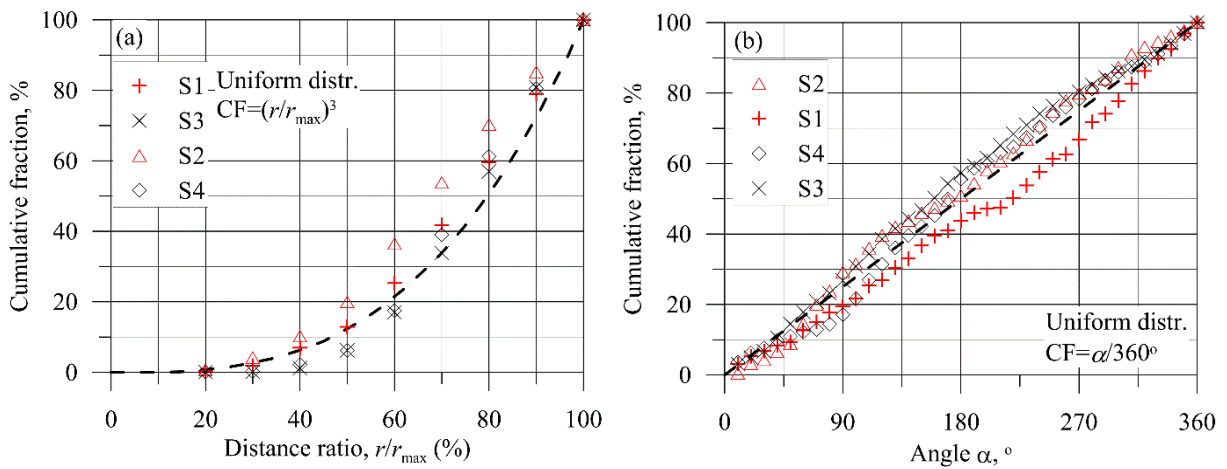

Figure 5: (a) Cumulative frequency of the distance ratio r/r$_{max}$; (b) Cumulative frequency for azimuthal angle a. Note that the dot lines indicate the theoretical curve of uniform distribution in a unit sphere and the corresponding theoretical equations are also given in the figure.

Figure 5a shows that the cumulative fraction of distance ratio is very close to the theoretical one for both conditions. It confirms that most parts of the specimen have been intruded with mercury. There is slightly variation, as for condition 1, the measured distribution is slightly higher than the theoretical one when distance ratio is between 40% to 90%. This indicates that mercury is more concentrated in this range. However, for condition 2, when distance ratio is between 30% to 60%, the measured distribution is below the theoretical, which indicates comparative smaller amount of mercury penetrates into this range when comparing with condition 1. This indicates that for specimens with larger void ratio (condition 1), mercury penetrates slightly deeper than for the specimens with a lower void ratio. The corresponding cumulative volume fraction with distance ratio is given in Figure S4, it shows similar behaviour as shown in Figure 5a. For condition 1, cumulative volume fraction at smaller distance ratio, in range between 20% to 40%, is also larger than that in condition 2. This indicates that mercury in condition 1 penetrates deeper than that in condition 2.

Figure 5b shows that the cumulative fraction of azimuthal angle $\alpha$ is very close to the theoretical one for both conditions. It confirms that for the mercury penetrates into the specimen from all directions and without a preferential direction. For specimen 1, the measured distribution is below the theoretical one, it is mainly due to the plateau when azimuthal angle between 180° to 225°. But the physical reason for this phenomenon is not very clear, it might due to the inhomogenity of specimen.

### 4.3 Shape of the mercury volumes

Beyond the simple description of "near spherical" and "non-spherical", the shape of each isolated mercury volume can be characterized by its flatness and anisotropy, both defined as the function of eigenvalue of the covariance matrix for moments of inertia (see supplementary material). As shown in supplementary material, the ratio of square root of eigenvalue is proportional to the length of the main axis of the mercury. Based on the definition of flatness and anisotropy defined in section 3.2, the values of 0 or 1 for flatness represent the length of the smallest main axis of mercury is equal to 0 or equal to the length of the medium main axis of mercury, respectively. For anisotropy value equal to 0 or 1 which indicate the length of the smallest main axis of mercury is equal to that of the largest main axis of mercury or equal to 0.

Before discussing the shape of isolated mercury volumes, it is relevant to assess how shape changes in a "flatness - anisotropy" space. The largest eigenvalue of covariance matrix for moments of inertia for all shapes shown in Figure 6 is set equal to 1. As the length of the largest main axis of mercury must be larger than that of medium main axis, in the flatness-anisotropy space, all data can only distribute in top half of the space. For the data along the diagonal line, based on the definition of flatness and anisotropy, the length of the medium main axis of mercury equal to that of largest main axis. So, from point (0, 1) to point (1,0), the shape of mercury gradually changes from sphere (at point 0,1) to a 2D circle (at point 1,0). In addition, when increasing anisotropy at constant flatness (e.g. flatness=1), the shape changes from a sphere to an approximately elongated rod and finally approaching to a line. In contrast, when increasing flatness at constant anisotropy (e.g. anisotropy=0.5), the shape changes from a disc-like ellipsoid into a rod-like ellipsoid.

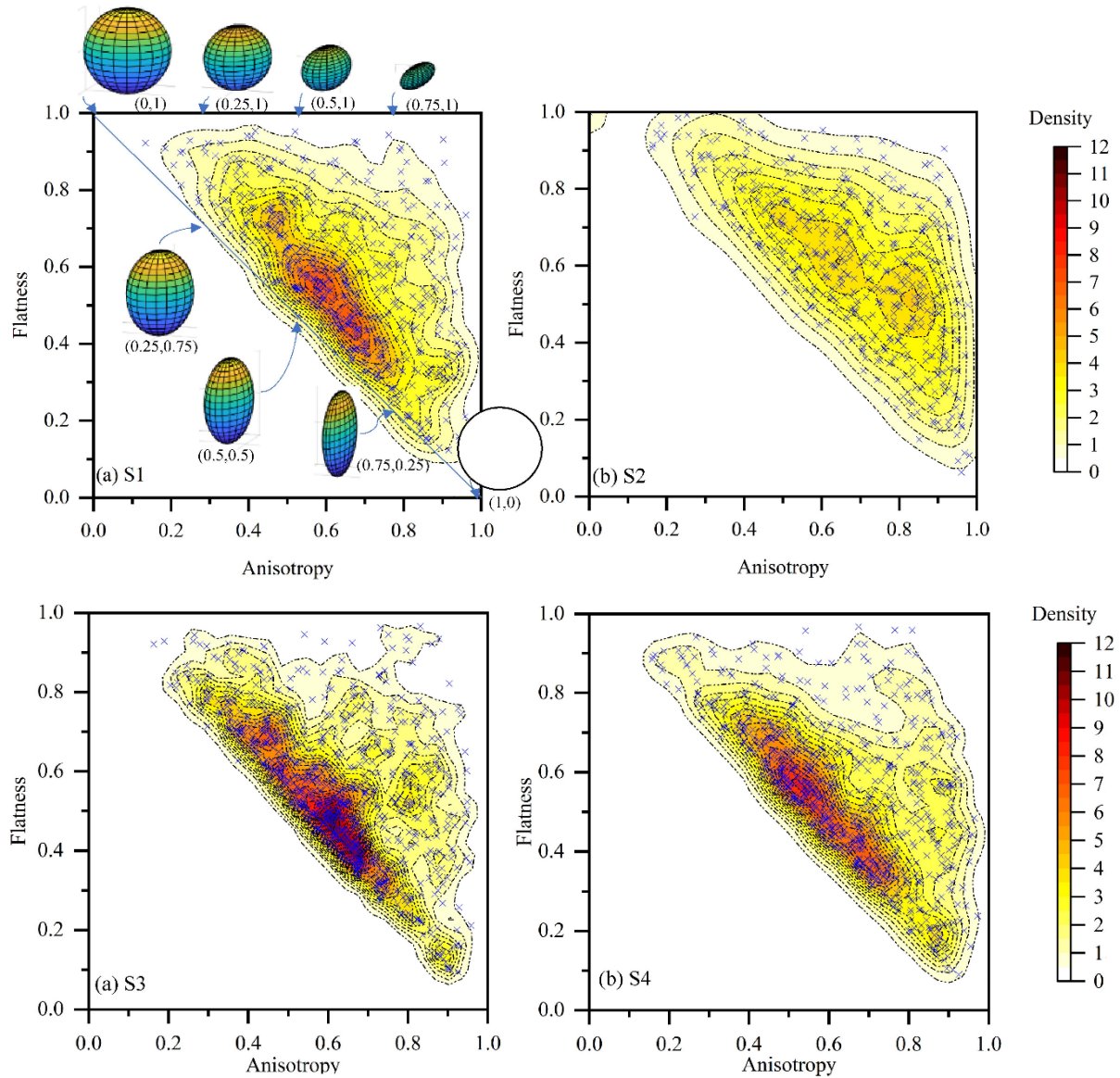

Figure 6: 2D kernel density estimation for the anisotropy and flatness of mercury of condition 1 ($w_o$=21%, $e_o$=1.0): (a) S1, and (b) S2; condition 2 ($w_o$=13.2%, $e_o$=0.8): (c) S3, and (d) S4.

The probability density function for the characteristics of mercury shape was estimated by 2D kernel density estimation, which is a non-parametric technique for probability density functions (Silverman, 1986). The contribution of each data point is smoothed out from a single point into a region of vicinity. The 2D kernel density estimation about the distribution of shape characteristics is presented in Figure 6a and 6b for specimens S1 and S2 (condition 1) and in Figure 6c and 6d for specimens S3 and S4 (condition 2)

The first observation is that the distribution of shape is quite similar for specimens S1 and S2 and for specimens S3 and S4, which indicates a satisfactory repeatability of results. The only noticeable difference is the maximum density of about 9 for specimen 1 as opposed to 5 for S2. A lower density value reflects a more uniform distribution of shapes, in the "flatness-anisotropy" space. The distribution of shapes for condition 2 (Figure 6c and 6d) is quite different from that of condition 1. In condition 2, the data is highly concentrated along the diagonal line, with anisotropy between 0.3 to 0.7, and flatness between 0.3 to 0.7. According to the results presented in Figure 6, there exist different shapes of mercury volumes, which differs from the general assumption of cylindrical pore geometry required to turn mercury pressure into pore diameter by Young-Laplace equation.

### 4.4 Ink-bottle geometry

Figure 7a shows the 3D structure of the largest mercury in non-spherical shape in specimen S3. It shows quite a complicated structure with pores of different size, here referred to as bodies and throats, a throat being a smaller pore connecting to larger pores called bodies.

A pore geometry with throats and bodies, commonly referred to as "ink-bottle" shape, cannot be adequately captured from mercury intrusion porosimetry. Indeed, the large body will only be intruded if the pressure is high enough for the mercury to first intrude the throat. Consequently, the volume of throat plus body is associated to the diameter of the throat. As a results, the volume of small pores is overestimated and volume of larger pores is underestimated (Giesche, 2006).

In this study, in order to characterize the ink-bottle geometry, a skeletonization process was carried out, which resulted in the mercury being represented by a network of tubes of varying thickness (see Figure 7b). The thickness at each point along the tube was representative of the pore size (Alim et al., 2017; Delerue et al., 1999) and the skeleton of tubes consisted of segments connected by nodes, as shown in Figure 7b and 7c. In particular, Figure 7b illustrates how colouring is used in conjunction of actual segment size to emphasize their thickness (Avizo, 2018). Bodies correspond to the thickest segments while throats are the thinnest ones.

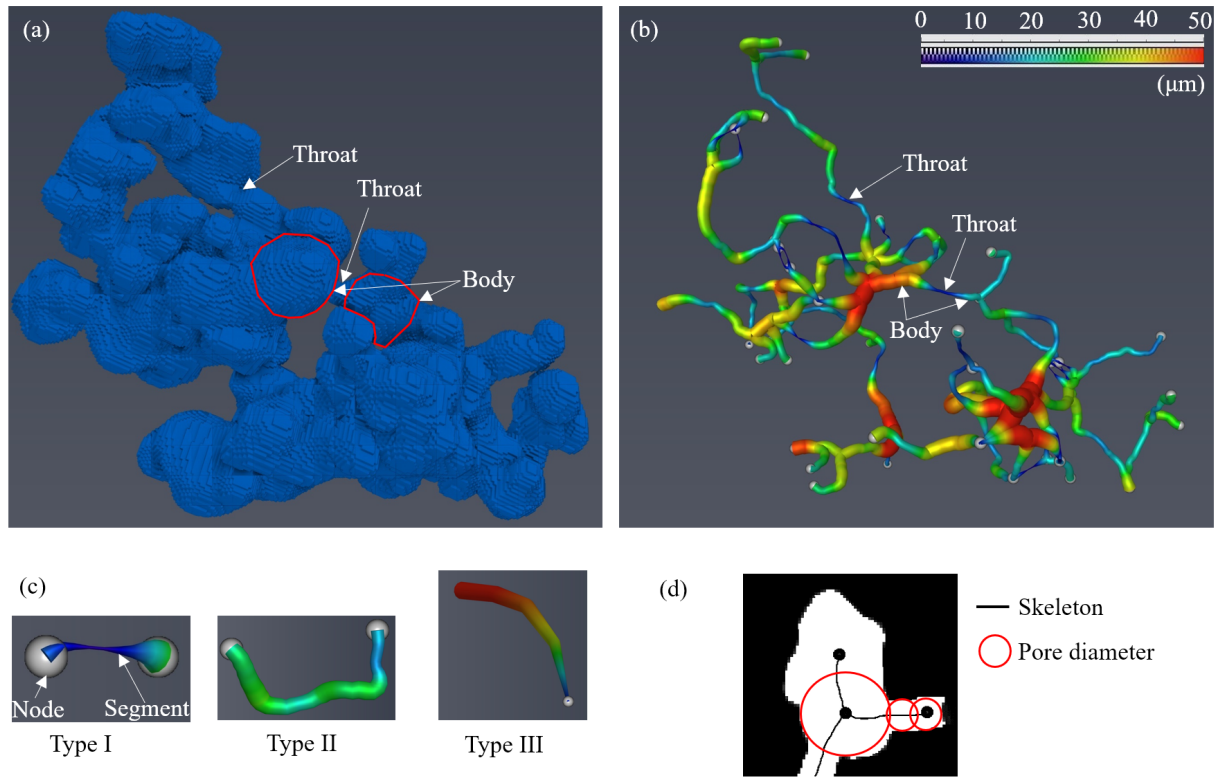

Figure 7: (a) 3D view and (b) skeletonization of the largest mercury in non-spherical shape in S3; (c) schematic for the three types of segments; (d) illustration of how the 2D pore space is mapped by a network of tubes with varying thickness. Type I: thickness of segment at the two nodes is larger than in the middle; Type II: thickness of segment is larger at its middle part; Type III: thickness of segment increases or decreases monotonically along the segment.

The skeletonization process reveals three main types of segment, depending on the change in thickness along the segment (see Figure 7c).Type I corresponds to segment whose thickness is larger at the two ends (nodes) than in the middle part of the segment; for type II, the thickness in the middle part of the segment is larger than that at the two nodes and, finally, for Type III, the thickness of the segment increases or decreases monotonically along the segment.

The relative proportions of type I, II and III segments were analysed and results are reported in Table 2. It can be seen that types I and II segments, that have an ink-bottle shape, represent more than 80% of all segments for all specimens analysed. Such high proportion suggests that the analysis of pore size distribution of these specimens will be subjected to an ink-bottle effect, the magnitude of which depends on the difference in size between bodies and throats.

Table 2: Relative proportion of each segment types for all specimens.

| Specimens | | Percentage of segment type, % | | |
|---|---|---|---|---|
| | | Type I | Type II | Type III |
| Condition 1 | S1 | 61.8 | 27.0 | 11.2 |
| | S2 | 60.3 | 23.8 | 15.9 |
| Condition 2 | S3 | 55.9 | 29.4 | 14.7 |
| | S4 | 63.6 | 25.0 | 11.4 |

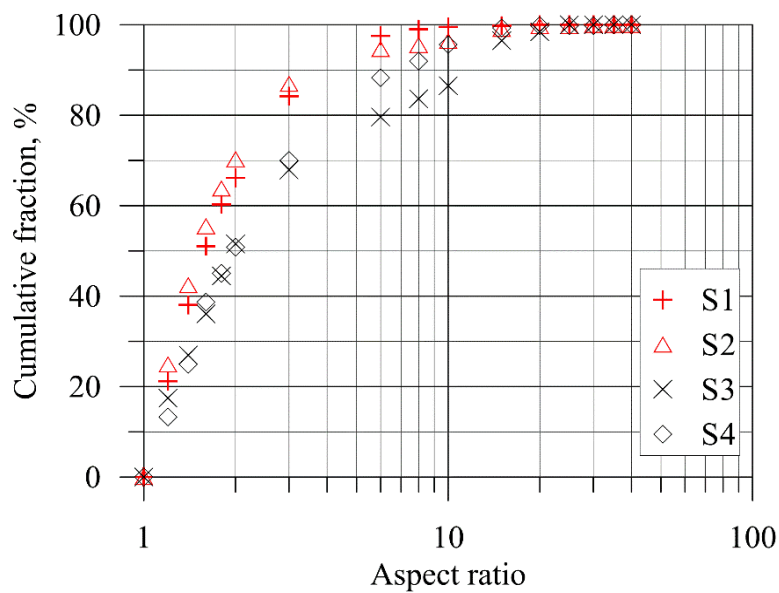

Figure 8: Relationship between aspect ratio and accumulative percent for the 4 specimens.

The aspect ratio is here defined as the ratio between the maximum and minimum thickness along the segment. The aspect ratio was computed for all type I and II segments and cumulative distributions were produced (see Figure 8). In interpreting the results of Figure 8, it is important to consider that Wardlaw & McKellar (1981) showed that for a pore structure with inner pore-size and throat ratio (i.e. aspect ratio) smaller than 3, the amount of mercury entrapped in the pore structure after releasing of pressure was relatively small. In the context of the present study, this implies that a significant ink-bottle effect can be expected for values of aspect ratio larger than 3. Figure 8 shows that the proportions of pores with an aspect ratio larger than 3 is less than 20% for S1 and S2, and 30% for S3 and S4. Overall, specimens in condition 2 have a larger proportion of pores with an aspect ratio over 3.

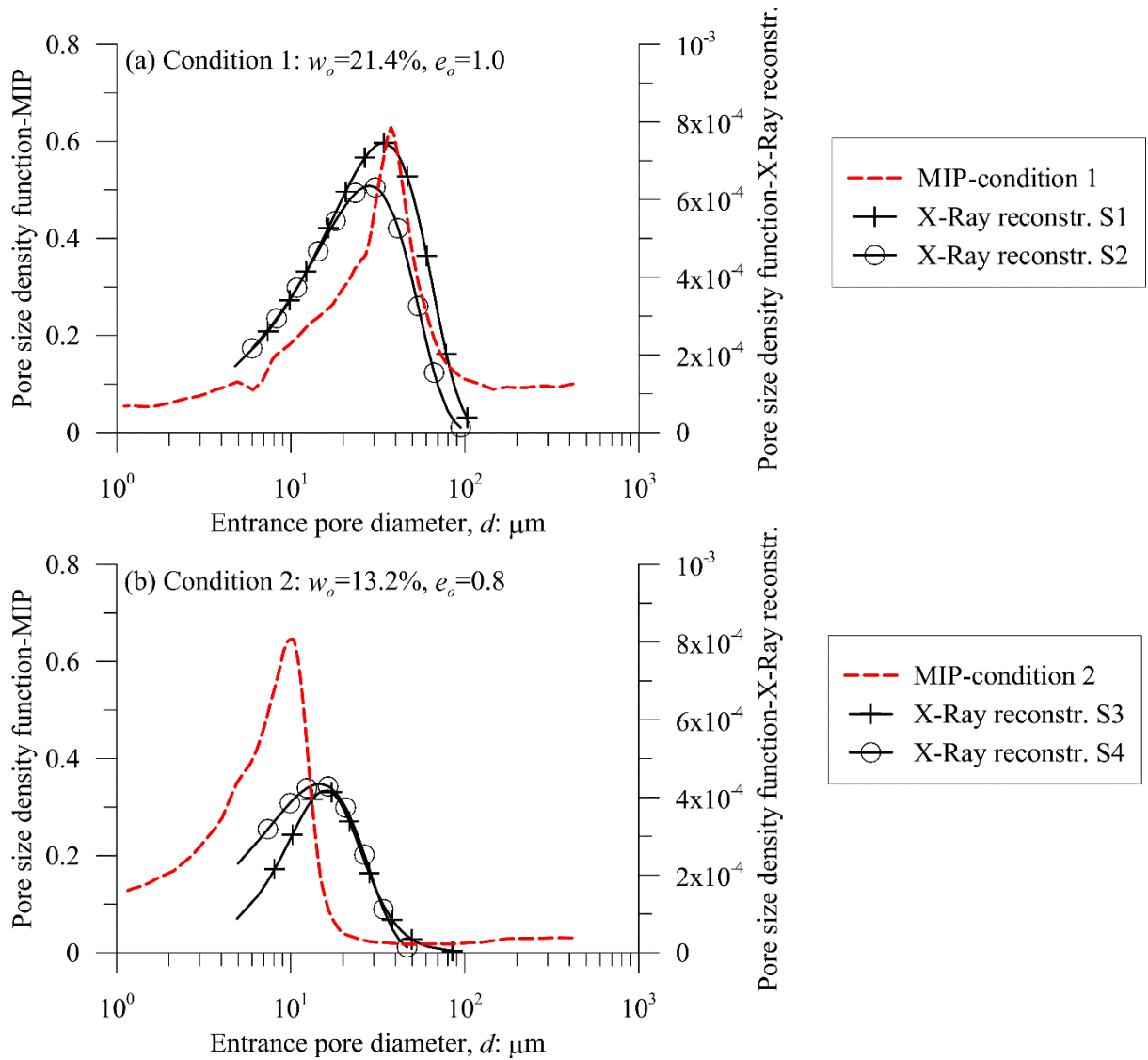

Figure 9: Comparison between the pore size distribution obtained from mercury intrusion porosimetry tests and X-μCT data: (a) condition 1; and (b) condition 2. Voxel size on images obtained by X Ray u CT is equal to 5 μm.

The question that now arises is whether a proportion of 20 to 30% of pores with a pronounced ink-bottle shape is sufficient to affect the results of mercury intrusion porosimetry. In order to provide elements of answer, the pore size distribution of the specimens was reconstructed from the X-μCT processed skeleton (i.e Figure 7b) and compared to those obtained by MIP. For each segment, i.e. pore, the entrance pore size was taken as the minimum thickness of the segment and the corresponding volume at that entrance pore size was taken as the volume of the segment. The pore size distributions reconstructed from X-μCT data are given in Figure 9. Figure 9 shows that the reconstructed PSD based X-μCT are almost the same for specimens with same condition. The height of peak for the reconstructed PSDs is much smaller than that obtained from MIP tests, which might be caused by the extrusion of some superficial mercury during unloading (release of mercury during sample preparation for X-μCT tests) and that fact that some mercury volumes are too small to be detected by X-μCT. A similar phenomenon was observed by Zeng et al. (2019), with just around 4% to 5% of total volume of intruded mercury detected by X-CT. For condition 1, as shown in Figure 9a, the dominant entrance pore sizes obtained from MIP (around 37 μm) and X-μCT (average around 32 μm) are very close to each other, with a slightly shift of dominant entrance pore size obtained from X-μCT to smaller value. For condition 1, the pores with aspect ratio larger than 3 were more or less uniformly distributed in the specimen (see Figure 10a), and such pores only represent around 20% of total amount of pores. Consequently, it is expected that the ink-bottle effect is only moderate. As present by Giesche (2006), entrapped mercury in artificial glass pore structures, after depressurization, the entrapped amount of mercury decreases as the aspect ratio decreasing. It is believed that some mercury was lost from the specimen after the porosimetry (some isolated mercury near spherical shapes were found outside the specimen, sealed in the epoxy but not shown in the paper for conciseness). The loss of mercury from the largest pores causes a slight shift of PSD obtained by X-μCT. This effect is less pronounced for condition 2, which has a stronger ink bottle effect and the capacity to better retain the mercury upon depressurization.

For condition 2, as shown in Figure 9b, there is an obvious difference of dominant entrance pore size: 10 μm (obtained from MIP) against 18 μm (obtained from X-μCT). Such difference might be correlated to the ink-bottle geometry. It is accepted that, upon unloading, mercury can snap-off at throat/body connection with some mercury remaining entrapped in the large pore. When interpreting MIP test results a large pore is allocated the size of its throat (what is referred to as the entrance pore diameter) and the combined volume of throat and body. In contrast, in the X-μCT analyses, where the ink-bottle ratio is pronounced, the mercury is mostly entrapped in the large pore with no mercury in the throat. Consequently, the large pore is assigned the diameter of the pore body and its volume. This difference leads to the shift of dominant entrance pore size in X-μCT to larger diameter values and lower intruded volumes, compared to the MIP results. As shown in Figure 10b, for condition 2, the mercury volumes with aspect ratio larger than 3 tend to be more concentrated on the out parts of the specimens, with $(r/r_{max})^{1.5}$ between 0.25 ($r/r_{max}$ around 0.40) and 0.65 ($r/r_{max}$ around 0.75). While for condition 1, Figure 11a, the mercury volumes with aspect ratio larger than 3 tend distributed more or less even when $(r/r_{max})^{1.5}$ is smaller than 0.6 ($r/r_{max}$ around 0.71). It is believed that a more concentrated distribution of mercury in superficial range of specimen would lead to the inner pore bodies become harder to be penetrated, and also harder to be released.

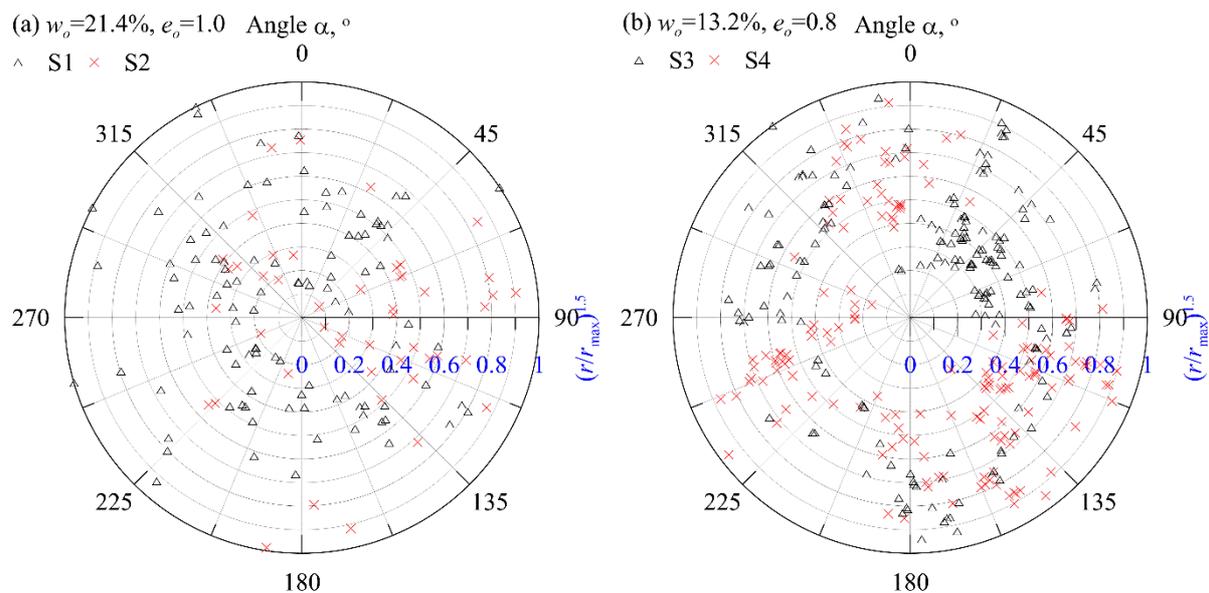

Figure 10: Polar plot of angle $\alpha$ and distance ratio $r/r_{max}$ for mercury with aspect ratio larger than 3: a) specimens S1, S2 (condition 1, $w_o$=21%, $e_o$=1.0); b) specimens S3, S4 (condition 2, $w_o$=13.2%, $e_0$=0.8).

# 5. Conclusion

To better investigate the pore morphology of compacted Maryland clay, X-Ray microtomography (X-µCT) tests were carried out on samples previously submitted to mercury intrusion porosimetry (MIP), i.e. that still contained an amount of entrapped mercury resulting from ink-bottle geometry. Two conditions were studied: condition 1 with initial water content, $w_o$, equal to 21%, and initial void ratio, $e_o$, equal to 1; condition 2 with $w_o$ equal to 13.2% and $e_o$ equal to 0.8. The analysis of the features of mercury detected by X-µCT suggested that macropores were almost uniformly distributed in specimen. For the two conditions studied in this paper, when the applied intrusion pressure was around the pressure corresponding to dominant entrance pore size of macropore, mercury could penetrate into most parts of specimens from all directions. For both conditions, most of mercury volumes were in non-spherical shape, and with aspect ratio smaller than three, like a twisted tube with an approximate constant elliptical cross section. The amount and distribution of mercury with aspect ratio larger than three produced a pronounced effect on pore size distribution (PSD) results obtained from MIP, which led to PSD obtained from MIP shift to smaller entrance pore size. This difference might be also due to the specific experimental protocol used in tests, i.e. only the large pore bodies with high aspect ratio were imaged in X-µCT, due to the extrusion of mercury during the process of depressurization and subsequent sample preparation for X-µCT. But entire pore space was accessible in MIP. The effect of intrusion pressures on the penetration of mercury in compacted clayey specimen will be studied in future.

# 6. Acknowledgements

The authors express their gratitude to the financial support from Natural Science Foundation of China (No. 52008355, No. 52078432, and No. 52168066), Science and Technology Program of Sichuan (No. 2021YJ0036 and No. 2019YFSY0015), and Southwest Jiaotong University (No. 2682020CX68).

# 7. References

AFNOR, 1999. Qualité des sols. Méthodes chimiques. Détermination de la ca- pacité d'échange cationique (CEC) et des cations extractives. French Stand. X 31-130, Paris.

Ahmed, S., Lovell, C.W.J., Diamond, S., 1974. Pore size and strength of compacted clay. J. Geotech. Eng. Div. 100, 407–425.

Alim, K., Parsa, S., Weitz, D.A., Brenner, M.P., 2017. Local Pore Size Correlations Determine Flow Distributions in Porous Media. Phys. Rev. Lett. 119, 1–5.

Andersson, L., Herring, A., Schlüter, S., Wildenschild, D., 2018. Defining a novel pore-body to pore-throat "Morphological Aspect Ratio" that scales with residual non-wetting phase capillary trapping in porous media. Adv. Water Resour. 122, 251–262.

Attari, A., McNally, C., Richardson, M.G., 2016. A combined SEM-Calorimetric approach for assessing hydration and porosity development in GGBS concrete. Cem. Concr. Compos. 68, 46–56.

Avizo, 2018. Avizo 3D software use's guide.

Bruchon, J.F., Pereira, J.M., Vandamme, M., Lenoir, N., Delage, P., Bornert, M., 2013. Full 3D investigation and characterisation of capillary collapse of a loose unsaturated sand using X-ray CT. Granul. Matter 15, 783–800.

Cases, J.M., Bérend, I., François, M., 1997. Mechanism of adsorption and desorption of water vapor by homoionic montmorillonite: 3. The Mg (super 2+), Ca (super 2+), and Ba (super 3+) exchanged forms. Clays Clay Miner. 45, 8–22.

Charalampidou, E.M., Hall, S.A., Stanchits, S., Lewis, H., Viggiani, G., 2011. Characterization of shear and compaction bands in a porous sandstone deformed under triaxial compression. Tectonophysics 503, 8–17.

Cnudde, V., Boone, M.N., 2013. High-resolution X-ray computed tomography in geosciences: A review of the current technology and applications. Earth-Science Rev. 123, 1–17.

Collins, K., Mcgown, A., 1974. The Form and Function of Microfabric Features in a Variety of Natural Soils. Geotechnique 24, 223–254.

Delage, P., 2010. A microstructure approach to the sensitivity and compressibility of some eastern Canada sensitive clays. Geotechnique 60, 353–368.

Delage, P., Audiguier, M., Cui, Y.-J., Howat, M.D., 1996. Microstructure of a compacted silt. Can. Geotech. J. 33, 150–158.

Delage, P., Lefebvre, G., 1984. Study of the structure of a sensitive Champlain clay and of its evolution during consolidation. Can. Geotech. J. 21, 21–35.

Delage, P., Marcial, D., Cui, Y.J., Ruiz, X., 2006. Ageing effects in a compacted bentonite: a microstructure approach. Géotechnique 56, 291–304.

Delage, P., Tessier, D., Audiguier, M.M., 1982. Use of the cryoscan apparatus for observation of freeze-fractured planes of a sensitive Quebec clay in scanning electron microscopy. Can. Geotech. J. 19, 111–114.

Delerue, J.F., Perrier, E., Yu, Z.Y., Velde, B., 1999. New algorithms in 3D image analysis and their application to the measurement of a spatialized pore size distribution in soils. Phys. Chem. Earth, Part A Solid Earth Geod. 24, 639–644.

Diamond, S., 1970. Pore size distributions in clays. Clays Clay Miner. 18, 7–23.

Fityus, S.G., Smith, D.W., 2004. The development of a residual soil profile from a mudstone in a temperate climate. Eng. Geol. 74, 39–56.

Fityus, S.G., Smith, D.W., Allman, M.A., 2004. Expansive Soil Test Site Near Newcastle. J. Geotech. Geoenvironmental Eng. 130, 686–695.

Fonseca, J., O'Sullivan, C., Coop, M.R., Lee, P.D., 2012. Non-invasive characterization of particle morphology of natural sands. Soils Found. 52, 712–722.

Garcia-Bengochea, I., Lovell, C.W., Altschaeffl, A.G., 1979. Pore Distribution and Permeability of Silty Clays. J. Geotech. Eng. Div. 105, 839–856.

Gerke, K.M., Sizonenko, T.O., Karsanina, M. V, Lavrukhin, E. V, Abashkin, V. V, Korost, D. V, 2020. Improving watershed-based pore-network extraction method using maximum inscribed ball pore-body positioning. Adv. Water Resour. 140, 103576.

Giesche, H., 2006. Mercury porosimetry: A general (practical) overview. Part. Part. Syst. Charact. 23, 9–19.

Gillott, J.E., 1973. Methods of sample preparation for microstructurel analysis of soil, in: Rutherford, G.K. (Ed.), Soil Microscopy Proceeding 4th International Working Meeting Soil Micromorphology. Limestone Press, Kingston, Ontario, Canada, pp. 143–164.

Iassonov, P., Gebrenegus, T., Tuller, M., 2009. Segmentation of X-ray computed tomography images of porous materials: A crucial step for characterization and quantitative analysis of pore structures. Water Resour. Res. 45, 1–12.

Lapierre, C., Leroueil, S., Locat, J., 1990. Mercury intrusion and permeability of Louiseville clay. Can. Geotech. J. 27, 761–773.

Le, T.X., Bornert, M., Aimedieu, P., Chabot, B., King, A., Tang, A.M., 2020. An experimental investigation on methane hydrate morphologies and pore habits in sandy sediment using synchrotron X-ray computed tomography. Mar. Pet. Geol. 122, 104646.

Liu, X., Zhang, C., Yuan, S., Fityus, S., Sloan, S.W., Buzzi, O., 2016. Effect of High Temperature on Mineralogy, Microstructure, Shear Stiffness and Tensile Strength of Two Australian Mudstones. Rock Mech. Rock Eng. 49, 3513–3524.

Liu, X.F., Buzzi, O., Yuan, S., Mendes, J., Fityus, S., 2016. Multi-scale characterization of the retention and shrinkage behaviour of four Australian clayey soils. Can. Geotech. J. 53, 854–870.

Massat, L., Cuisinier, O., Bihannic, I., Claret, F., Pelletier, M., Masrouri, F., Gaboreau, S., 2016. Swelling pressure development and inter-aggregate porosity evolution upon hydration of a compacted swelling clay. Appl. Clay Sci. 124–125, 197–210.

Monroy, R., Zdravkovic, L., Ridley, A., 2010. Evolution of microstructure in compacted London Clay during wetting and loading. Géotechnique 60, 105–119.

Moro, F., Böhni, H., 2002. Ink-bottle effect in mercury intrusion porosimetry of cement-based materials. J. Colloid Interface Sci. 246, 135–149.

Ng, C.W.W., Sadeghi, H., Jafarzadeh, F., Sadeghi, M., Zhou, C., Baghbanrezvan, S., 2020. Effect of microstructure on shear strength and dilatancy of unsaturated loess at high suctions. Can. Geotech. J. 57, 221–235.

Otalvaro, I.F., Neto, M.P.C., Delage, P., Caicedo, B., 2016. Relationship between soil structure and water retention properties in a residual compacted soil. Eng. Geol. 205, 73–80.

Otsu, N., 1979. A threshold selection method from gray-level histograms. IEEE Trans. Syst. Man. Cybern. 9, 62–66.

Ravikovitch, P.I., Neimark, A. V., 2002. Experimental confirmation of different mechanisms of evaporation from ink-bottle type pores: Equilibrium, pore blocking, and cavitation. Langmuir 18, 9830–9837.

Romero, E., 2013. A microstructural insight into compacted clayey soils and their hydraulic properties. Eng. Geol. 165, 3–19.

Romero, E., Della Vecchia, G., Jommi, C., 2011. An insight into the water retention properties of compacted clayey soils. Géotechnique 61, 313–328.

Romero, E., Simms, P.H., 2008. Microstructure investigation in unsaturated soils: A review with special attention to contribution of mercury intrusion porosimetry and environmental scanning electron microscopy. Geotech. Geol. Eng. 26, 705–727.

Saba, S., Barnichon, J.D., Cui, Y.J., Tang, A.M., Delage, P., 2014. Microstructure and anisotropic swelling behaviour of compacted bentonite/sand mixture. J. Rock Mech. Geotech. Eng. 6, 126–132.

Schluter, S., Sheppard, A., Brown, K., Wildenschild, D., 2014. Image processing of multiphase images obtained via X-ray microtomography:A review. Water Resour. Res. 50, 3615–3639.

Seiphoori, A., Ferrari, A., Laloui, L., 2014. Water retention behaviour and microstructural evolution of MX-80 bentonite during wetting and drying cycles. Géotechnique 64, 721–734.

Silin, D., Patzek, T., 2006. Pore space morphology analysis using maximal inscribed spheres. Phys. A Stat. Mech. its Appl. 371, 336–360.

Silverman, B.W., 1986. Density estimation for statistics and data analysis. Springer Science + Business Media B.V.

Sufian, A., Russell, A.R., 2013. Microstructural pore changes and energy dissipation in Gosford sandstone during pre-failure loading using X-ray CT. Int. J. Rock Mech. Min. Sci. 57, 119–131.

Tanino, Y., Blunt, M.J., 2012. Capillary trapping in sandstones and carbonates: Dependence on pore structure. Water Resour. Res. 48, 1–13.

Tovey, N.K., Wong, K.Y., 1973. The preparation of soils and other geological materials for the scanning electron microscope, in: Leenheer, L. De (Ed.), Proceedings of the International Symposium on Soil Structure. Swedish Geotechnical Society, Gothenburg, Sweden, pp. 176–183.

Wardlaw, N.C., McKellar, M., 1981. Mercury porosimetry and the interpretation of pore geometry in sedimentary rocks and artificial models. Powder Technol. 127–143.

Wong, H.S., Head, M.K., Buenfeld, N.R., 2006. Pore segmentation of cement-based materials from backscattered electron images. Cem. Concr. Res. 36, 1083–1090.

Yuan, S., Liu, X., Romero, E., Delage, P., Buzzi, O., 2020. Discussion on the separation of macropores and micropores in a compacted expansive clay. Géotechnique Lett. 10, 454–460.

Yuan, S., Liu, X., Sloan, S.W., Buzzi, O.P., 2016. Multi-scale characterization of swelling behaviour of compacted Maryland clay. Acta Geotech. 11, 789–804.

Yuan, S.Y., Buzzi, O., Liu, X.F., Vaunat, J., 2019a. Swelling behaviour of compacted Maryland clay under different boundary conditions. Géotechnique 69, 514–525.

Yuan, S.Y., Liu, X.F., Buzzi, O., 2021a. A microstructural perspective on soil collapse. Géotechnique 71, 132–140.

Yuan, S.Y., Liu, X.F., Buzzi, O., 2019b. Effects of soil structure on the permeability of saturated Maryland clay. Géotechnique 69, 72–78.

Yuan, S.Y., Liu, X.L., Buzzi, O., 2021b. Technical aspects on mercury intrusion porosimetry for clays. Environ. Geotech. 8, 255–263.

Zeng, Q., Chen, S., Yang, P., Peng, Y., Wang, J., Zhou, C., Wang, Z., Yan, D., 2020. Reassessment of mercury intrusion porosimetry for characterizing the pore structure of cement-based porous materials by monitoring the mercury entrapments with X-ray computed tomography. Cem. Concr. Compos. 113, 103726.

Zeng, Q., Wang, X., Yang, P., Wang, J., Zhou, C., 2019. Tracing mercury entrapment in porous cement paste after mercury intrusion test by X-ray computed tomography and implications for pore structure characterization. Mater. Charact. 151, 203–215.

Zimmie, T.F., Almaleh, L.J., 1976. Shrinkage of Soil Specimens During Preparation for Porosimetry Tests, in: Soil Specimen Preparation for Laboratory Testing, ASTM STP 599. American Society for Testing and Materials, pp. 202–215.